\documentclass[aps,11pt,nofootinbib,groupedaddress,preprintnumbers,superscriptaddress]{revtex4}
\usepackage[dvipdfmx]{graphicx}
\usepackage{amssymb,amsfonts,amsmath,bm,color,multirow,cases,empheq,hyperref}
\usepackage{mathrsfs}

\usepackage{enumerate}
\usepackage{ulem}
\usepackage{afterpage}
\hypersetup{%
 setpagesize=false,
 bookmarksnumbered=true,%
 bookmarksopen=true,%
 colorlinks=true,%
 linkcolor=blue,%
 citecolor=red}
\begin{document}

\title{
Stable circular orbits in Kaluza-Klein black hole spacetimes
}

\author{Shinya Tomizawa}
\email{tomizawa@toyota-ti.ac.jp}
\affiliation{Mathematical Physics Laboratory, Toyota Technological Institute, Nagoya 468-8511, Japan}

\author{Takahisa Igata}
\email{igata@post.kek.jp}
\affiliation{KEK Theory Center, Institute of Particle and Nuclear Studies, High Energy Accelerator Research Organization, Tsukuba 305-0801, Japan}

\preprint{TTI-MATHPHYS-4}
\preprint{KEK-TH-2308} 
\preprint{KEK-Cosmo-0274}

\begin{abstract} 
Reducing motion of particles to a two-dimensional potential problem, we show that there are stable circular orbits around a squashed Kaluza-Klein black hole with a spherical horizon and multi--Kaluza-Klein black holes with two spherical horizons in five dimensions.
For a single horizon, we show analytically that the radius of an innermost stable circular orbit monotonically depends on 
the size of an extra dimension. 
For two horizons, the radius of an innermost stable circular orbit depends on the separation between two black holes besides the size of an extra dimension.
More precisely, the set of the stationary points of the potential is composed of two branches.
For a large separation, stable circular orbits exist on the two branches regardless of the size of an extra dimension, and in particular, on one branch, the set of stable circular orbits is connected for the small extra dimension but has two disconnected parts for the large extra dimension.
For a small separation, only on one branch it exists, and the radius of an innermost stable circular orbit monotonically increases with an extra-dimension size. 
\end{abstract}

\date{\today}
\maketitle

\section{Introduction}

Higher-dimensional black holes have played an important role in understanding basic properties of fundamental theories, such as string theory. 
A number of interesting solutions of higher-dimensional black holes have been discovered so far, revealing much richer structure of their solution space than that of four-dimensional black holes, and we are naturally led to address the question of how to classify them~\cite{Emparan:2008eg}. 
However, since our real, observable world is macroscopically four-dimensional, extra dimensions have to be compactified in realistic, classical spacetime models~\cite{Kaluza:1921tu,Klein:1926tv}. 
Therefore, it is of great interest to consider higher-dimensional Kaluza-Klein black hole spacetimes, which look like four-dimensional ones at least at large distances. 
The studies on such Kaluza-Klein black holes may also help us to get some insights into the major open problem of how to compactify and stabilize extra dimensions in string theory. 
The simplest types of Kaluza-Klein black holes are direct products of four dimensions and extra dimensions. 
However, nontrivial, interesting classes of Kaluza-Klein black hole solutions can be obtained by a squashing transformation from the same class of noncompactified black hole solutions, such as asymptotically flat rotating black hole solutions with equal angular momenta, which are often called squashed Kaluza-Klein black holes~\cite{Tomizawa:2011mc}.

\medskip
The studies on stable bound orbits for particles moving around black holes is one of the interesting issues because it significantly depends on the spacetime dimensions, the topologies of black holes and the types of theories whether stable bound orbits exist or what the radius of an innermost stable circular orbit is. 
For instance, in a four-dimensional Schwarzschild background, such stable bound orbits exist for massive particles, and in contrast, in higher-dimensional Schwarzschild backgrounds they do not exist~\cite{Tangherlini:1963bw}. 
Moreover, as for the rotating black holes in five dimensions, in a Myers-Perry background with a spherical horizon~\cite{Myers:1986un}, any stable bound orbits do not exist~\cite{,Page:2006ka,Frolov:2003en,Frolov:2006pe}, whereas in a black ring background with the horizon topology of $S^1\times S^2$ and black lens backgrounds with horizon topologies of lens spaces $L(n;1)$ ($n:$ natural number) they exist~\cite{Igata:2010ye,Igata:2013be,Tomizawa:2019egx,Tomizawa:2020mvw}, and the shapes and number of the existence regions of stable bound orbits are quite different according to the horizon topologies. 
As is discussed in Refs.~\cite{Nakashi:2019mvs,Igata:2020vlx} for multi--black holes with two spherical horizons, 
in four dimensions~\cite{Nakashi:2019mvs}, stable circular orbits always exist from near-horizon to infinity, in five dimensions~\cite{Igata:2020vlx}, they exist only when the separation of two black holes is large enough, and in higher dimensions, they cannot exist regardless of the separation~\cite{Igata:2020vlx}.

\medskip
In our previous paper~\cite{Igata:2021wwj}, we studied the motion of massive particles in a five-dimensional spacetime with a compactified extra-dimensional space where a black hole is localized in the direction of extra-dimensional space, which is referred to as a caged black hole. 
We showed the existence of circular orbits and their stability for both massive and massless particles. 
In this paper, we study the motion of massive particles around a squashed Kaluza-Klein black hole with a compactified extra-dimensional space where a black hole extends. 
%where a black hole is expanding to the whole extra-dimensional space.
In our analysis, we consider motion of particles as a two-dimensional potential problem, focusing on the static charged black hole solutions with a single horizon and two horizons in the five-dimensional Einstein-Maxwell theory, which were constructed in Ref.~\cite{Ishihara:2006iv}. 
More precisely, a problem of whether there exist stable circular orbits for particles can be replaced with a simple problem of whether the two-dimensional effective potential has a local minimum. 
First, we analytically show the existence of stable circular orbits for a static charged black hole with a single horizon, which has a mass parameter (saturating a BPS bound) and an extra-dimension parameter. 
Near the horizon, the spacetime behaves like a five-dimensional black hole, whereas at infinity, it effectively behaves like a four-dimensional spacetime. 
Therefore, as can be easily expected, stable circular orbits do not exist near the horizon as the five-dimensional Schwarzschild black hole~\cite{Tangherlini:1963bw} but exist at infinity as the four-dimensional Schwarzschild black hole. 
Next, we numerically clarify the existence of stable circular orbits for static multi--charged black holes with two horizons, which have a separation parameter between two horizons, besides (equal) mass parameters of the black holes and the extra-dimension parameter. 
Moreover, we see that the existence region of stable circular orbits significantly changes on the separation of the black holes and the size of the extra dimension.

\medskip
The rest of the paper is composed as follows: 
In the following Sec.~\ref{sec:kkbh}, we briefly review the static solution of multi--Kaluza-Klein black holes in the five-dimensional Einstein-Maxwell theory (the five-dimensional minimal supergravity). 
In Sec.~\ref{sec:formalism}, we explain our formalism to find stable circular orbits. 
In Sec.~\ref{sec:SBO}, we show that there are stable circular orbits for a single black hole and two black holes.
In Sec.~\ref{sec:summary}, we summarize and discuss our results.

\section{Review of Kaluza-Klein black holes}\label{sec:kkbh}
For later use, we briefly review the static solutions of multi--Kaluza-Klein black holes in the five-dimensional Einstein-Maxwell theory~\cite{Ishihara:2006iv}. 
As is shown in the Appendix, these solutions can be obtained from the supersymmetric solutions in the five-dimensional minimal supergravity~\cite{Gauntlett:2002nw}. 
Note that this theory is a five-dimensional Einstein-Maxwell-Chern-Simons theory with a certain special coupling constant, but the Chern-Simons term vanishes for such a static charged case.

The metic and the gauge potential of the static solutions of multi--Kaluza-Klein black holes are written, respectively, as
\begin{eqnarray}
ds^2&=&-L^{-2}dt^2+L\left[ H^{-1}(d\zeta+\omega)^2+Hdx^idx^i  \right], \label{eq:metric}\\
     A&=&\frac{\sqrt{3}}{2} L^{-1}dt,
\end{eqnarray}
where $dx^idx^i=dx^2+dy^2+dz^2$ is the metric on three-dimensional Euclidean space ${\mathbb E}^3$, and $\zeta$ is a periodic coordinate with the range $0 \le \zeta \le 2\pi l$, where $l$ corresponds to the radius of the compactified extra dimension.
The functions $H$ and $L$ are the harmonic functions on $\mathbb{E}^3$ with point sources at ${\bm r}={\bm r}_i$ $[{\bm r}=(x,y,z), {\bm r}_i=(x_i,y_i,z_i), r_i=|{\bm r}-{\bm r}_i|]$,
\begin{eqnarray}
H=1+\sum_{i=1}^n\frac{N_i}{r_i},\quad L=1+\sum_{i=1}^n\frac{M_i}{r_i},
\end{eqnarray}
and the one-form $\omega=\omega_idx^i$ is given by
\begin{eqnarray}
\omega=\sum_{i=1}^n\frac{z-z_i}{r_i}\frac{(x-x_i)dy-(y-y_i)dx}{(x-x_i)^2+(y-y_i)^2}. 
\end{eqnarray}
Let us introduce the spherical coordinates $(r,\theta,\varphi)$ defined by 
\begin{eqnarray}
(x,y,z)=(r\sin\theta\cos\varphi,r\sin\theta\sin\varphi,r\cos\varphi),
\end{eqnarray}
where $r>0$, $0 \le \theta\le \pi$, $0 \le \varphi \le 2\pi$. 
A point source with $M_i > 0$ and $N_i = 0$ or with $M_i < 0$ or $N_i < 0$ corresponds to a naked singularity. 
Furthermore, a point source with $M_i = 0$ and $N_i>0$ corresponds to a Gross-Perry-Sorkin monopole with a nut charge $N_i$, in which case $N_i=l/2$ must be required from regularity at ${\bm r}={\bm r}_i$. 
A point source with $M_i> 0$ and $N_i> 0$ denotes a black hole with mass $M_i$, where $N_i=ln_i/2$ ($n_i$: natural numbers) is required from regularity on the horizon, whose topology is the lens space $L(n_i;1)=S^3/{\mathbb Z}_{n_i}$ because the induced metric on the $i$-th point source ${\bm r}={\bm r}_i$ is written as
\begin{eqnarray}
ds^2=\frac{lM_in_i}{2}\left[\left(\frac{d\psi }{n_i}+\cos \theta d\varphi\right)^2+(d\theta^2+\sin^2\theta d\varphi^2)\right],
\end{eqnarray}
where $0\le \psi:=2\zeta/l\le 4\pi$. 
For $n_i=1$, this coincides with the metric on $S^3$ written in the Euler angles, and for $n_i\ge 2$, this coincides with the metric on the lens space $L(n_i;1)=S^3/{\mathbb Z}_{n_i}$. At infinity $r\to\infty$, the metric~(\ref{eq:metric}) behaves as
\begin{eqnarray}
ds^2\simeq -dt^2+dr^2+r^2(d\theta^2+\sin^2\theta d\varphi^2)+\left[d\zeta+\sum_{i=1}^nN_i \cos\theta d\varphi \right]^2,
\end{eqnarray}
which has the structure of $S^1$ fiber bundle over the four-dimensional Minkowski spacetime.
Thus, it turns out that Eq.~(\ref{eq:metric}) describes the Kaluza-Klein black hole which looks five-dimensional near the horizon but four-dimensional at large distances.

In what follows, to find the stable circular orbits for test particles moving around such a Kaluza-Klein black hole, we consider only the case of $M_i,N_i>0$ and $n_i=1\ (i=1,\ldots ,n)$, in which the horizon topology of each black hole is $S^3$. 
In the analysis below, we use the cylindrical coordinates $(\rho,\varphi,z)$ defined by $(x,y,z)=(\rho\cos\varphi,\rho\sin\varphi,z)$.

\section{Our formalism} \label{sec:formalism}

Our method to find stable circular orbits is based on the previous work~\cite{Tomizawa:2019egx}.
Now, we give the brief review as follows. 
We assume that all the black holes exist on the $z$-axis, i.e., $x_i=y_i=0$, 
which does not lose generality for the cases $n=1, 2$.
Then, the spacetime described by Eq.~(\ref{eq:metric}) has three Killing vector fields $(\xi_t,\xi_\varphi,\xi_\zeta):=(\partial/\partial t,\partial/\partial \varphi,\partial/\partial \zeta)$, 
and therefore, 
three scalars 
$(E,L_{\varphi},L_{\zeta}):=(-g_{\mu\nu}\xi_t^\mu u^\nu,g_{\mu\nu}\xi_\varphi^\mu u^\nu, g_{\mu\nu}\xi_\zeta^\mu u^\nu )$ are constants of motion along a geodesic with a tangent vector $u^\mu:=dx^\mu/d\lambda$ ($\lambda$: affine parameter), which correspond to the energy, the angular momentum around the $z$-axis and the momentum along $\partial/\partial \zeta$ of a particle, respectively.
Then, the normalization condition $g_{\mu\nu}u^\mu u^\nu=-\kappa$ is written as
\begin{eqnarray}
\frac{H}{L}(\dot \rho^2+\dot z^2)+U=E^2,
\label{eq:Hamiltonian2}
\end{eqnarray}
where $U$ is the effective potential for massive particles with unit mass ($\kappa=1$) and massless particles ($\kappa=0$)
and is simply written as
\begin{eqnarray}
U=\frac{\kappa}{L^2}+\frac{L_\varphi^2+(H^2\rho^2+\omega_\varphi^2 )L_\zeta^2-2\omega_\varphi L_\varphi L_\zeta}{HL^3\rho^2}.
\end{eqnarray}
Thus, because we can consider the motion of the massive %and massless 
particles in the two-dimensional potential $U(\rho,z)$, the allowed motion regions of particles are restricted to $U\le E^2$. 
At a stationary point such that $\nabla_iU=0$ and $U=E^2$, there exists a circular orbit of particles, whose stability is determined from the positivity of the determinant and trace of the Hesse matrix $(\nabla_i \nabla_j U)\ (i,j=\rho,z)$ of $U$, 
where $\nabla_i$ is the covariant derivative 
associated with the two-dimensional conformally flat metric 
\begin{eqnarray}
ds^2=\frac{H}{L}(d\rho^2+dz^2).
\end{eqnarray}
Namely, if the conditions $\mathrm{Tr}(\nabla_i \nabla_j U)>0$ and $\mathrm{det}(\nabla_i \nabla_j U)>0$ hold at a stationary point $\nabla_iU=0$ and $U=E^2$, the potential $U$ has a local minimum at the point, where particles move on a stable circular orbit. 
Because $\nabla_i\nabla_j U=\partial_i\partial_j U$ 
at the stationary points,
the conditions can be replaced with $\mathrm{Tr}(H_{ij})>0$ and $\mathrm{det}(H_{ij})>0$ for $(H_{ij}):=(\partial_j \partial_i U)$.

\medskip
Throughout this paper, for simplicity we consider the case of $L_\zeta=0$, 
where $U$ can be written as 
\begin{eqnarray}
U(\rho, z; L_\varphi^2)=\frac{\kappa}{L^2}+\frac{L_\varphi^2}{HL^3\rho^2}.
\end{eqnarray}
The set of such stationary points of the potential $U$ is determined by
\begin{eqnarray}
U_{\rho}&=&-2\kappa\frac{L_{\rho}}{L^3}-\frac{\rho(H_\rho L+3HL_\rho)+2HL}{H^2L^4\rho^3}L_\varphi^2=0,\label{eq:Ur}\\
U_{z}&=&-2\kappa\frac{L_{z}}{L^3}-\frac{H_z L+3HL_z}{H^2L^4\rho^2}L_\varphi^2=0,\label{eq:Uz}\\
U&=&E^2. \label{eq:UE}
\end{eqnarray}
From Eq.~(\ref{eq:Ur}), the squared angular momentum $L_{\varphi}^2$ is written as
\begin{eqnarray}
L_{\varphi}^2
=-\kappa\frac{2\rho^3H^2LL_\rho}{(H_\rho L+3HL_\rho)\rho+2HL}
=:L_{\varphi 0}^2(\rho,z), 
\label{eq:Lphi0}
\end{eqnarray}
and from this, Eqs.~(\ref{eq:Uz}) and (\ref{eq:UE}) are written, respectively, as
\begin{eqnarray}
U_{z}(\rho,z;L_{\varphi 0}^2)&=&-2\kappa\frac{2HL_z +H_\rho L_z \rho-H_zL_\rho \rho}{L^2(2HL+H_\rho L \rho+3H L_\rho \rho)}, \label{eq:Uz0}\\
E^2&=&U(\rho,z;L_{\varphi 0}^2)=\frac{\kappa}{L^2}-\frac{2\kappa\rho H L_\rho}{[(H_\rho L+3HL_\rho)\rho+2HL]L^2}=:E_{0}^2(\rho,z).
\label{eq:E0}
\end{eqnarray}

We define the set $\gamma_0$ of stationary points by
\begin{eqnarray}
\gamma_0=\{(\rho,z) \:\!|\:\! U_{z}(\rho,z;L_{\varphi 0}^2)=0            \} ,
\end{eqnarray}
which denotes certain curves on the two-dimensional $(\rho,z)$-space. 
From the relation between $E_0^2$ and $L_{\varphi 0}^2$,
\begin{eqnarray}
E^2_0=\frac{\kappa}{L^2}\left(1+\frac{L_{\varphi 0}^2}{\rho^2HL}\right),
\end{eqnarray}
we find that $E_0^2\ge 0$ is satisfied whenever $L_{\varphi 0}^2\ge 0$ is satisfied. 
We define the two-dimensional region $D$ on the $(\rho,z)$-space by 
\begin{eqnarray}
D=\{(\rho,z) \:\!|\:\! 
h_0>0, k_0>0, L_{\varphi 0}^2\ge 0               \}, 
\end{eqnarray}
where
\begin{align}
h_0(\rho, z)
&:=\mathrm{det}(H_{ij}(\rho, z; L_{\varphi 0}^2)),
\\
k_0(\rho, z)
&:=\mathrm{Tr}(H_{ij}(\rho, z; L_{\varphi 0}^2)).
\end{align}
To summarize, showing the existence of stable circular orbits can be reduced to finding the overlap of the curves $\gamma_0$ and the region $D$ on the two-dimensional $(\rho,z)$-space.

\medskip
Let us note that our formalism cannot be applied to massless particles ($\kappa=0$) because Eqs.~(\ref{eq:Lphi0})--(\ref{eq:E0}) vanish. 
However, within the framework of $\kappa=1$, we can obtain the results of $\kappa=0$ by observing the limit in which $E_0$ and $L_{\varphi 0}^2$ diverge, respectively, while keeping the ratio $L_{\varphi 0}/E_0$ finite.
This is why in the analysis below, we consider only the motion of massive particles.

\section{Stable circular orbits}\label{sec:SBO}
\subsection{Single black hole}
First, we show the existence of stable circular orbits for the static charged black hole with a single spherical horizon ($n=1$), which can be obtained by setting the parameters in Eq.~(\ref{eq:metric}) as $(M_1,N_1,n_1)=(M,N,1)$ and ${\bm r}_1=(0,0,0)$.
The set $\gamma_0$ of the stationary points is restricted on $z=0$ because $U_{z}(\rho,z;L_{\varphi 0}^2)=0$ is equivalent with $z=0$. 
Therefore, let us consider only the geodesic motion of particles on the $\rho$-axis $(z=0)$, where 
$E_{0}^2$, $L_{\varphi 0}^2$, 
$k_0$ and $h_0$
are reduced to, respectively, 
\begin{eqnarray}
\label{eq:E0s}
E_{0}^2(\rho, 0)&=&\frac{\rho^3(2\rho+M+N)}{(\rho+M)^2f_1(\rho)},\\
\label{eq:L0s}
L_{\varphi 0}^2(\rho,0)&=&\frac{2M(\rho+M)(\rho+N)^2}{f_1(\rho)},\\
\label{eq:k0s}
k_0(\rho, 0)
&=& \frac{2M[4\rho^4+2(M+2N)\rho^3+(M^2-MN+2N^2)\rho^2+(M+N) MN\rho+2M^2N^2] }{(\rho+M)^4(\rho+N)f_1(\rho)},\\
h_0(\rho, 0)&=& \frac{8M^2\rho g_1(\rho) }{(\rho+M)^6 f_1(\rho)^2},
\end{eqnarray}
where 
\begin{eqnarray}
&&f_1(\rho):=2\rho^2-(M-N)\rho-2MN,\\
&&g_1(\rho):=2\rho^3-2M\rho^2-M(M+9N)\rho-3MN(M+N).
\end{eqnarray}
The conditions
$E_0^2(\rho, 0) > 0$, $ L_{\varphi 0}^2(\rho, 0) > 0$ and $k_0(\rho, 0)>0$  
are equivalent with $f_1>0$, which can be rewritten as the range of $\rho$,
\begin{align}
\rho >\rho_1
:=\frac{M-N+\sqrt{(N+7M)^2-48M^2}}{4}.
\end{align}
On the other hand, the condition $h_0(\rho, 0)>0$ reduces to $g_1>0$, 
or equivalently, 
\begin{align}
\label{eq:rho2}
\rho>\rho_2(M, N),
\end{align}
where $\rho_2$ is the unique positive root of the cubic equation $g_1=0$. 
Furthermore, for any $M>0$ and $N>0$, 
the following inequalities hold:
\begin{align}
g_1(\rho_1)
=-4MN(M+N)+\frac{1}{2}(N^2-14MN-3M^2)\rho_1
<\frac{N^2}{2}(\rho_1-8M)
<0,
\end{align}
where in the last inequality, we have used the upper bound of $\rho_1$,
\begin{align}
\rho_1<\frac{M-N+\sqrt{(N+7M)^2}}{4}=2M.
\end{align}
As a result, the conditions $f_1>0$ and $g_1>0$ are equivalent to Eq.~\eqref{eq:rho2}.
This means that the region $\gamma_0\cap D$ is reduced to a semi-infinite line $\rho>\rho_2$ on $z=0$.
Hence, it turns out that $\rho_2$ 
corresponds to the radius of the innermost stable circular orbit.
 \begin{figure}[t]
\includegraphics[width=8cm]{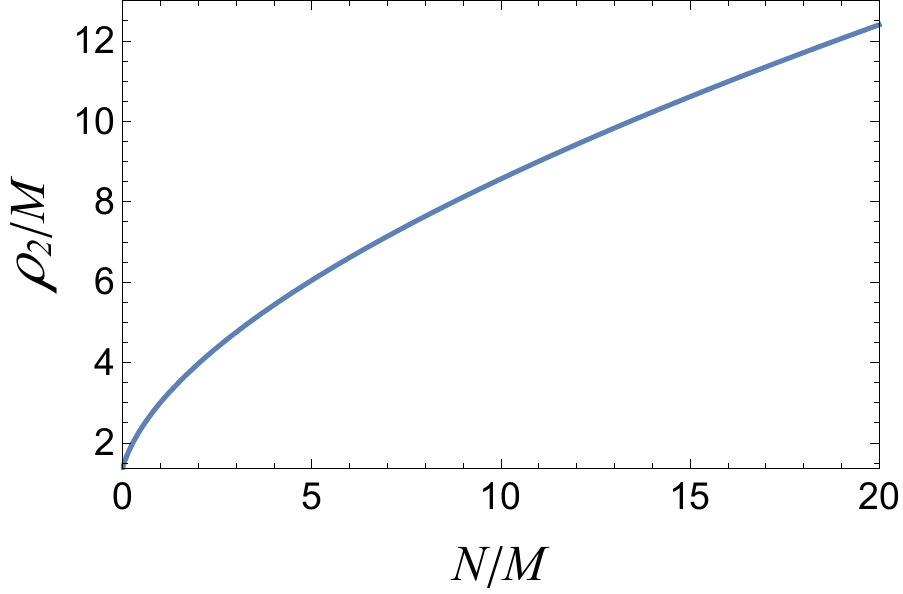}
\caption{The dependence of $\rho_2$ on $N/M$}
\label{fig:1b}
\end{figure}

As is shown in Fig.~\ref{fig:1b}, $\rho_2$ is a monotonically increasing function of $N/M$. 
In particular, in the small limit of the extra dimension size $N/M(=l/(2M))\to 0$, $\rho_2$ takes a minimum as $\rho_2/M \to (1+\sqrt{3})/2= 1.366...$. 
In the large limit of $N/M\to \infty$, we have $\rho_2\to\infty$, and stable circular orbits vanish because the limiting solution is a five-dimensional asymptotically flat black hole.

\subsection{Two black holes}

Next we consider the solution for $n=2$ in Eq.~(\ref{eq:metric}) with two spherical horizons at $z=\pm z_1$ ($z_1>0$) on the $z$-axis, which can be obtained by setting the parameters in Eq.~(\ref{eq:metric}) as $(M_1,N_1,n_1)=(M_2,N_2,n_2)=(M,N,1)$ and ${\bm r}_1=-{\bm r}_2=(0,0,z_1)$.

The set $\gamma_0$ of the stationary points is determined by
\begin{eqnarray}
U_{z}(\rho,z;L_{\varphi 0}^2)=0 ~ \Longleftrightarrow ~ (r_1^3+r_2^3)z+(r_1^3-r_2^3)z_1=0, \label{eq:Uz0}
\end{eqnarray}
which describes curves in the $(\rho,z)$-plane.
The roots of Eq.~(\ref{eq:Uz0}) have two branches, $\{(\rho,z)|z=0 \}$ and $\{(\rho,z)|(r_1^3+r_2^3)z+(r_1^3-r_2^3)z_1=0,z\not=0 \}$, which we call ``$z=0$ branch'' 
and ``$z\not=0$ branch'', respectively. 

\begin{figure}[t]
\centering
\includegraphics[width=6cm]{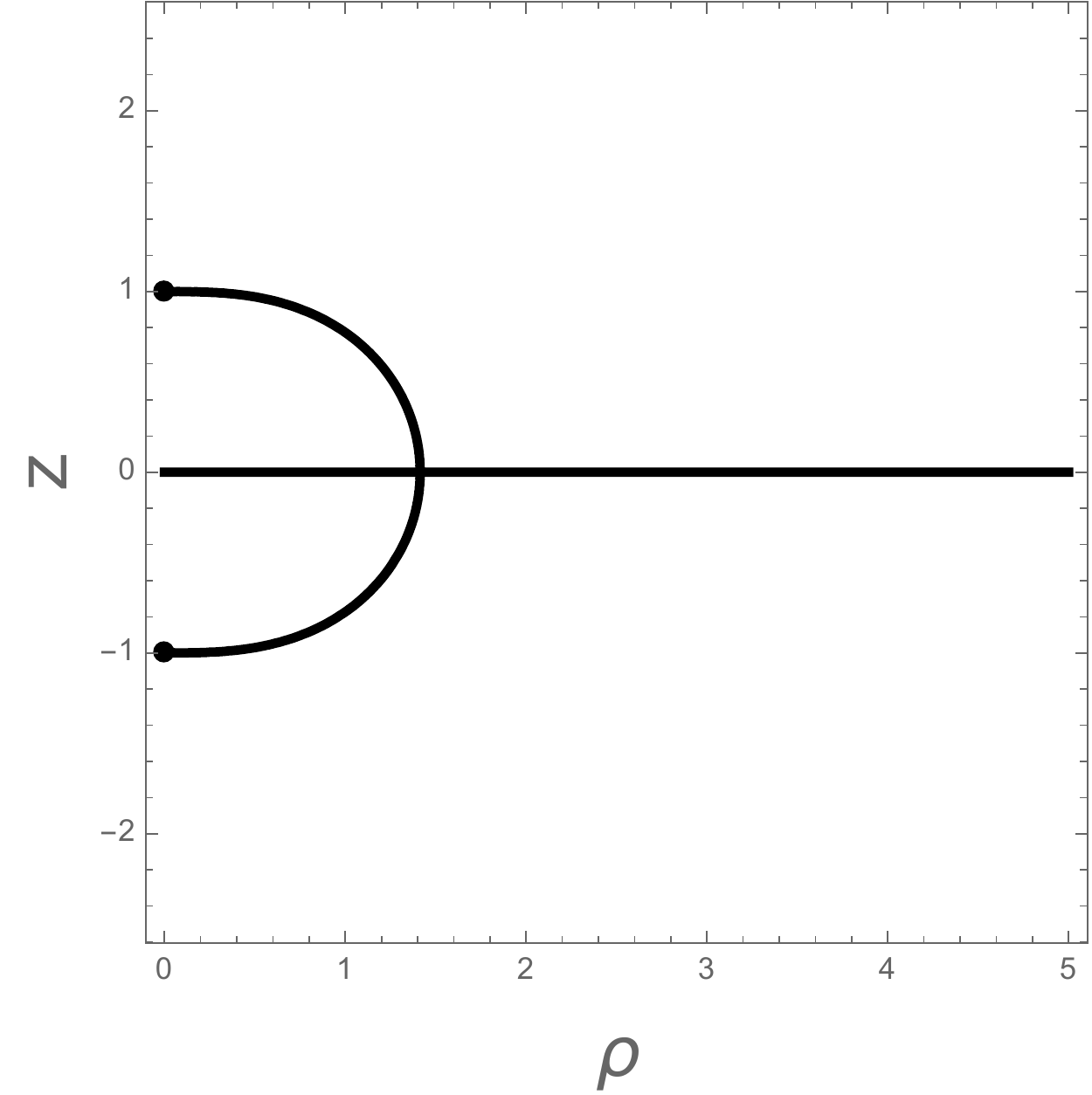}

\caption{Curves denote the set $\gamma_0$ of stationary points for $z_1=1$, which has two branches, the $z=0$ branch and the $z\not=0$ branch. Two black points at $(\rho,z)=(0,\pm 1)$, which are endpoints of the $z\not=0$ branch, correspond to two black holes. }

\label{fig:sp_1}
\end{figure}

Next let us denote the graph corresponding to the $z\not=0$ branch by a function $\rho=\rho(z)$. 
Equation~(\ref{eq:Uz0}) can be written as
\begin{eqnarray}
\frac{z}{z_1}=F(\rho,z):=-\frac{r_1^3-r_2^3}{r_1^3+r_2^3},
\end{eqnarray}
and hence we find that the function $\rho=\rho(z)$ exists in the range $0<|z|<z_1$. 
From 
$F(\sqrt{2}z_1,0)=0$
and 
$F(0,\pm z_1)=\pm 1$, 
the graph intersects with the $\rho$-axis at $\rho=\sqrt{2}z_1$ and with the $z$-axis at $z=\pm z_1$. 
From the derivative 
\begin{eqnarray}
\frac{d\rho}{dz}=-\frac{z_1F_z-1}{z_1F_\rho}
\left\{
\begin{array}{ll}
<0 & (0<z<z_1)\\
=0 &(z=0) \\
>0 & (-z_1<z<0)
\end{array}
\right.,
\end{eqnarray}
it turns out that $\rho(z)$ is monotonically decreasing for $z>0$ and monotonically increasing for $z<0$, 
which is illustrated in Fig.~\ref{fig:sp_1}.

We discuss whether the stationary orbit at a point on $\gamma_0$ are circular or not. It is useful to clarify our definition of the circular orbit in higher-dimensional spacetimes. If a particle orbit projected onto a preferred time slice (e.g., a Killing time slice) coincides 
with a closed orbit of an axial Killing vector, we call the particle orbit the circular orbit. 
From our assumption, we have $L_\zeta=g_{\mu\nu}\xi_\zeta^\mu u^\nu=g_{\zeta \zeta}(\dot\zeta+\omega_\varphi\dot\varphi)=0$. A particle staying at a point on 
the $z=0$ branch satisfies $\dot \zeta=0$ because $\omega_{\varphi}$ vanishes on $z=0$. Then, 
the particle orbit projected onto the static time slice coincides with a closed orbit of $\partial/\partial \varphi$, and thus, the stationary orbit of such a particle is circular. 
On the other hand, the stationary orbit of a particle staying at a point $(\rho(z_0), z_0)$ on 
the $z\neq 0$ branch is not always circular. 
It is circular if $\omega_\varphi(\rho(z_0), z_0)$ is a rational number, whereas 
it is noncircular if $\omega_\varphi(\rho(z_0), z_0)$ is an irrational number. 
Nevertheless, in what follows we call the stationary orbits on the $z\neq 0$ branch the circular orbits  because they are a dense set.

\begin{figure}[t]
\begin{tabular}{cc}
\begin{minipage}[t]{0.5\hsize}
\centering
\includegraphics[width=6cm]{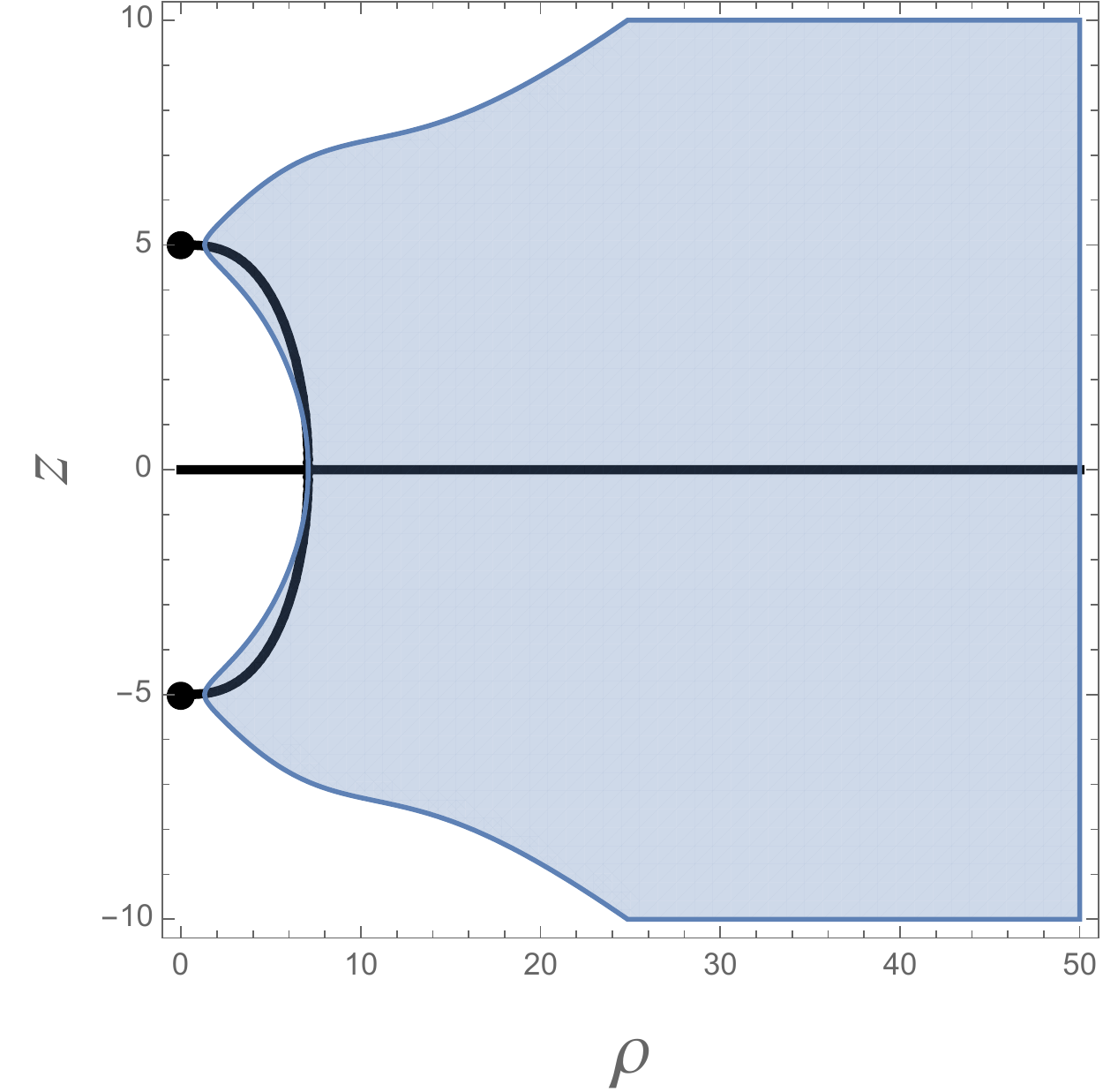}
~~~~
\caption{$M=1$, $z_1=5$, $l=0.05$
}
\label{fig:a5_1}
~~
\end{minipage}
& 
\begin{minipage}[t]{0.5\hsize}
\centering
\includegraphics[width=6cm]{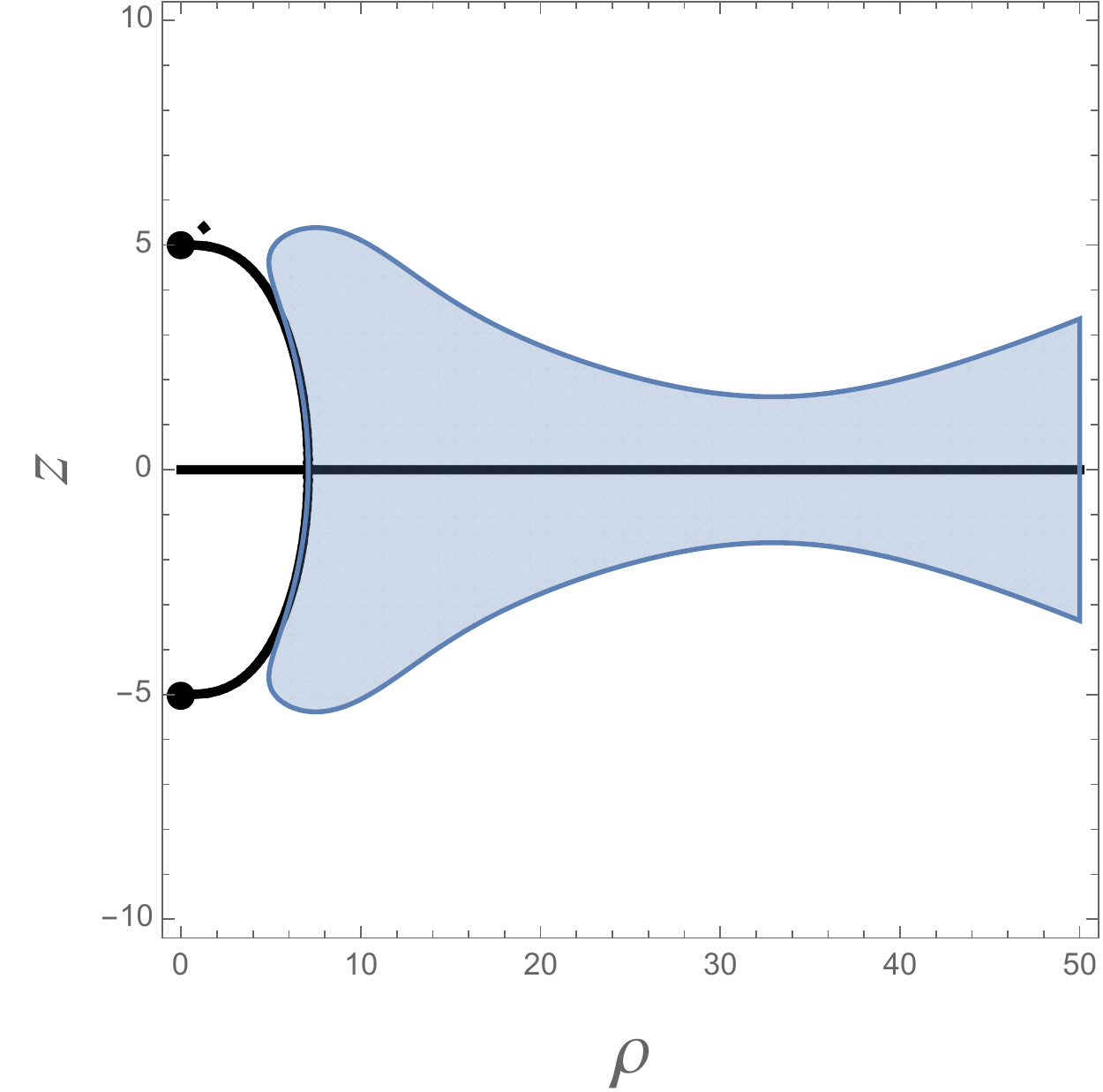}
~~~~
\caption{$M=1$, $z_1=5$, $l=150$}
\label{fig:a5_4_3}
\end{minipage} 
\\
 \begin{minipage}[t]{0.5\hsize}
 \centering
\includegraphics[width=6cm]{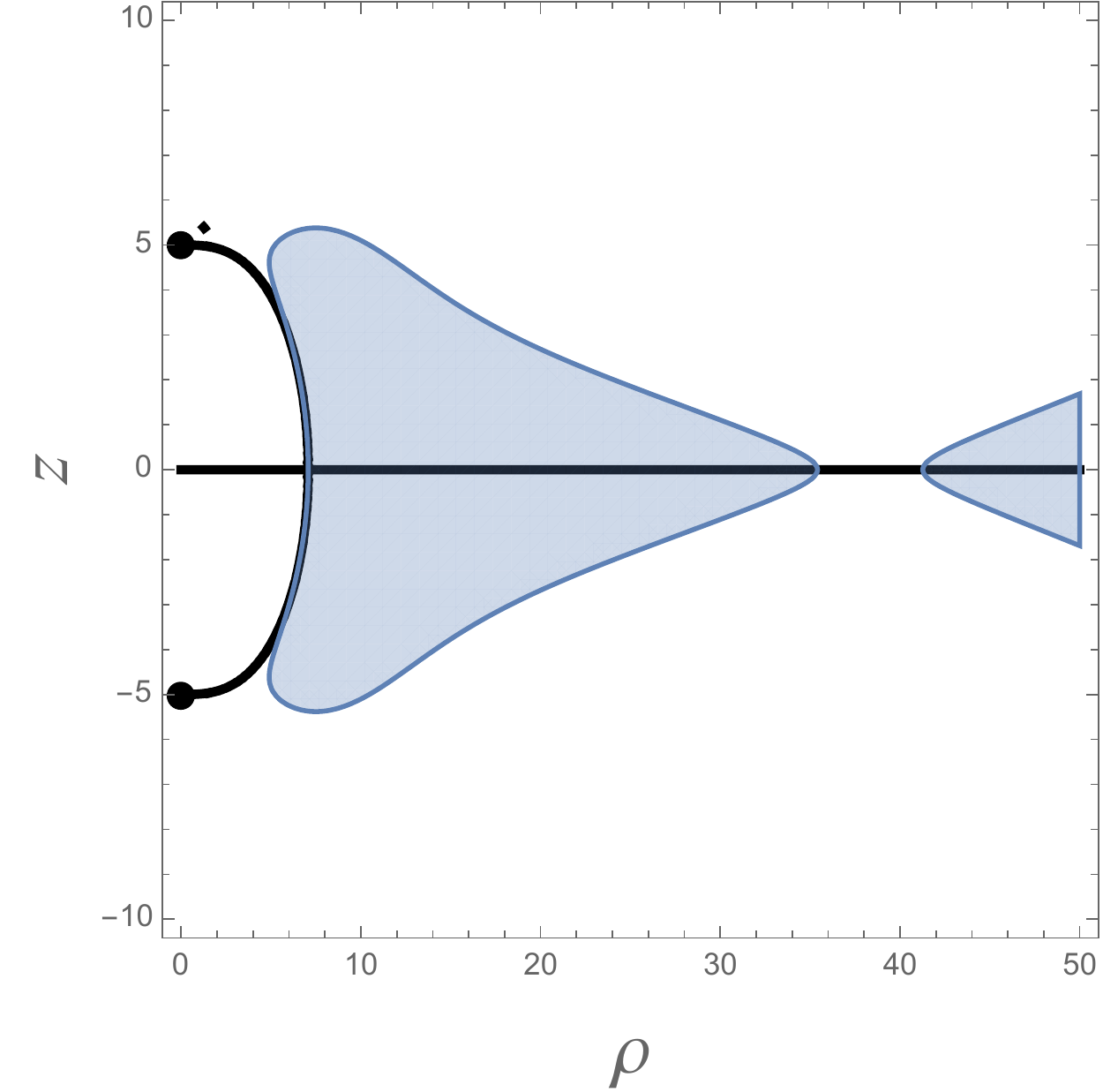}
~~~~
\caption{$M=1$, $z_1=5$, $l=200$
}
\label{fig:a5_4_4}
 \end{minipage} 
 &
\begin{minipage}[t]{0.5\hsize}
\centering
\includegraphics[width=6cm]{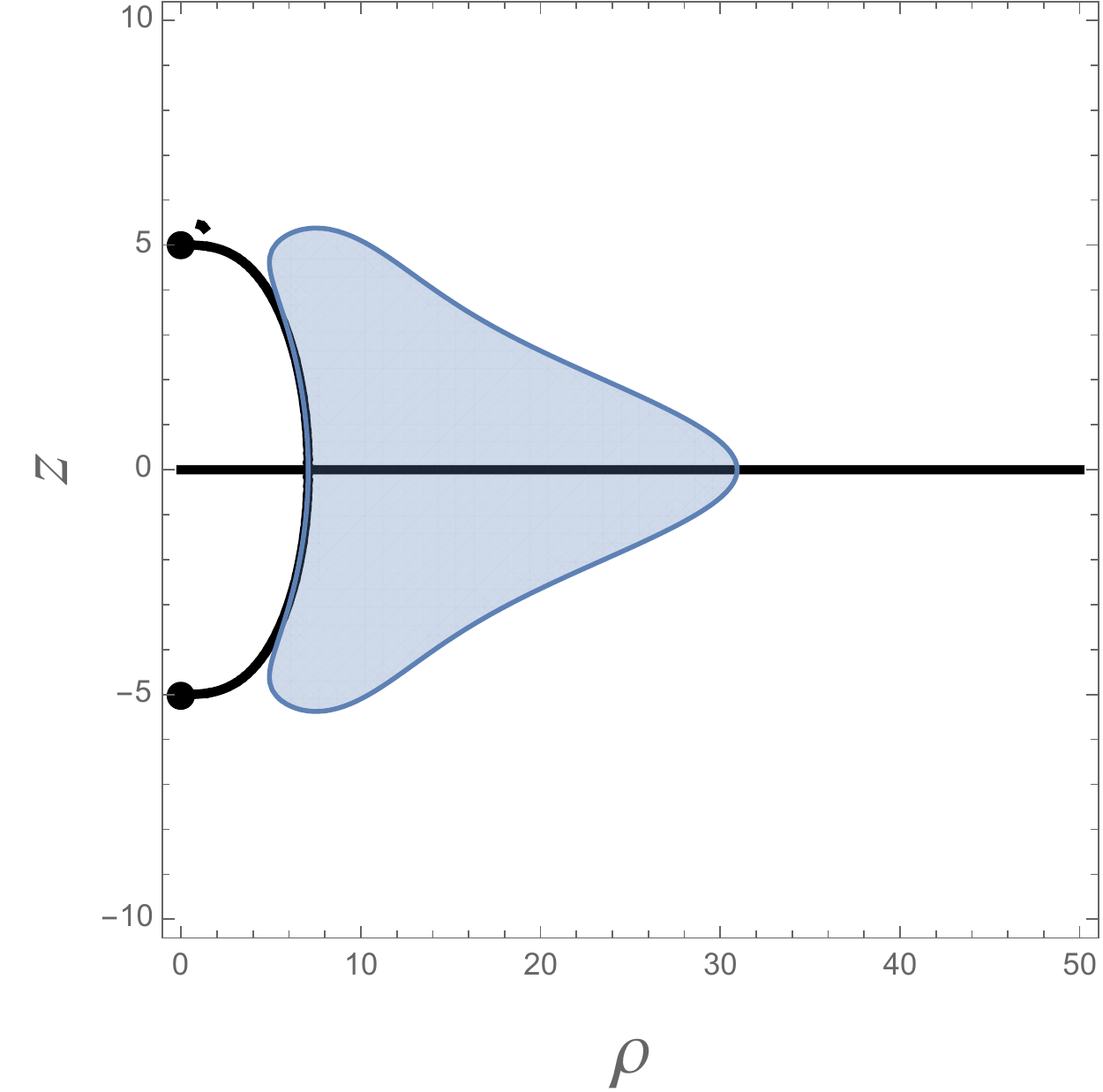}
~~~~
\caption{$M=1$, $z_1=5$, $l=250$
}
\label{fig:a5_4_5}
 \end{minipage} 

\end{tabular}
\end{figure}

\newpage

Figures~\ref{fig:a5_1}--\ref{fig:a5_4_5} show the dependence of the existence region of stable circular orbits on $l$ for a large separation under the parameter setting $(M,z_1)=(1,5)$ and $l=0.05, 150, 200, 250$, where 
 the black solid curves and the blue shaded regions denote $\gamma_0$ and $D$, respectively. 
On the $z=0$ branch, the stable circular orbits extend from the point $(\rho,z)=(\sqrt{2}z_1,0)$ to infinity $(\rho,z)=(\infty,0)$ for small $l$, whereas they are composed of two portions, a finite portion $\sqrt{2}z_1<\rho<\rho_{\mathrm{in}}$ and a semi-infinite portion $\rho_{\mathrm{out}}<\rho<\infty$ $(0<\sqrt{2}z_1<\rho_{\mathrm{in}}<\rho_{\mathrm{out}})$,  for large $l$. 
Moreover, in the limit $l\to \infty$, only a finite portion exists because the semi-infinite portion goes to infinity and vanishes, 
which is similar to the case of the single horizon. 
On the $z\not=0$ branch, the stable circular orbits extend from the points $(\rho,z)=(\rho(z_\mathrm{I}),\pm z_\mathrm{I})\ (0<z_\mathrm{I}<z_1)$ near the horizon to the point $(\rho,z)=(\sqrt{2}z_1,0)$ on the $\rho$-axis. 
Furthermore, it can be seen from these figures that as $l$ becomes small, $\rho(z_{\mathrm{I}})$ becomes smaller, namely, 
the radius of innermost stable circular orbits becomes smaller.

\begin{figure}[t]
\begin{tabular}{cc}
\begin{minipage}[t]{0.5\hsize}
\centering
\includegraphics[width=6cm]{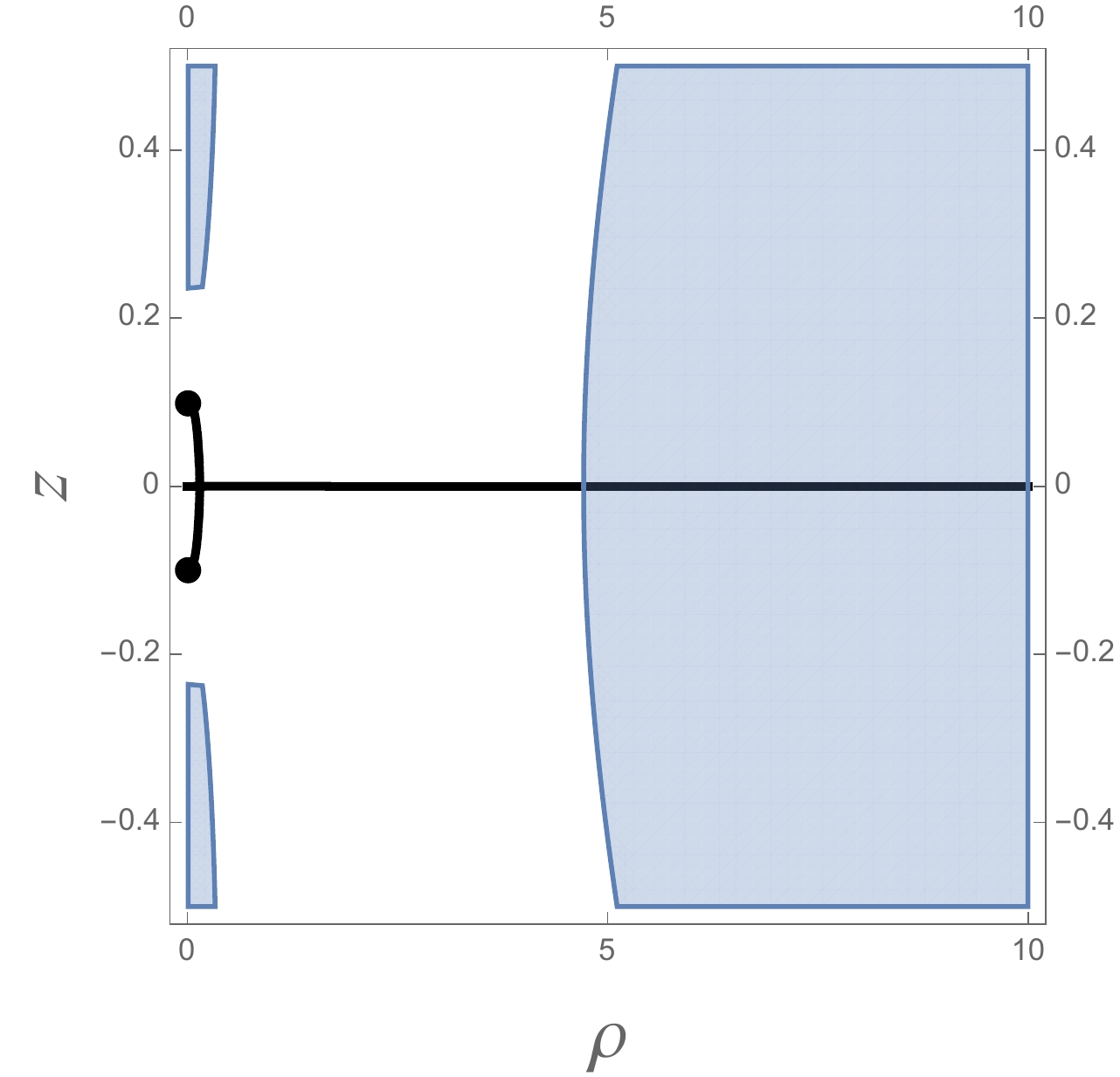}
~~~~
\caption{$M=1$, $z_1=0.1$, $l=1$
}
~~
\label{fig:a01_4}
 \end{minipage} &
 
\begin{minipage}[t]{0.5\hsize}
\centering
\includegraphics[width=6cm]{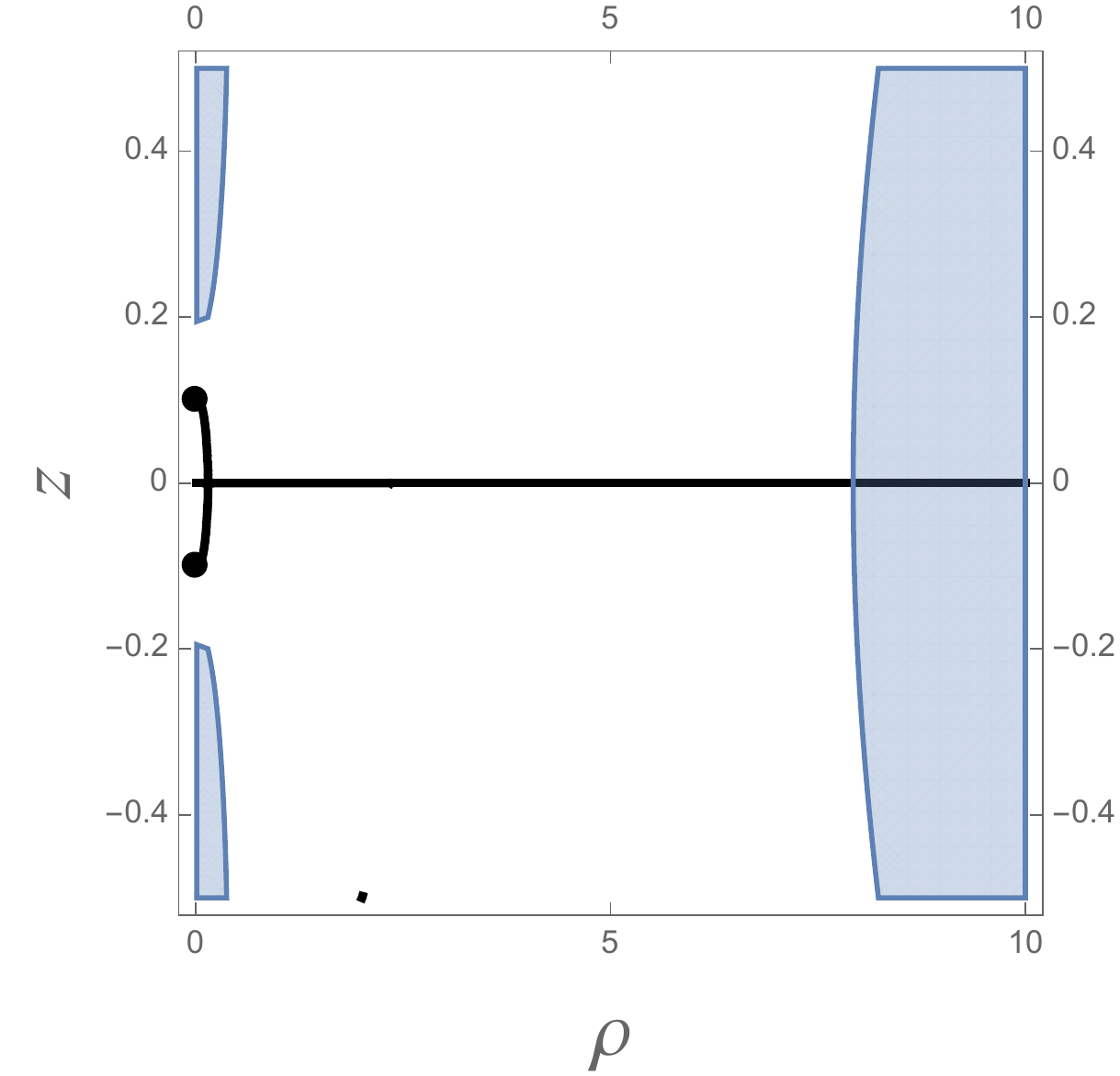}
~~~~
\caption{$M=1$, $z_1=0.1$, $l=4$
}
\label{fig:a01_4_4}
 \end{minipage} \\
 
 \medskip

\begin{minipage}[t]{0.5\hsize}
\centering
\includegraphics[width=6cm]{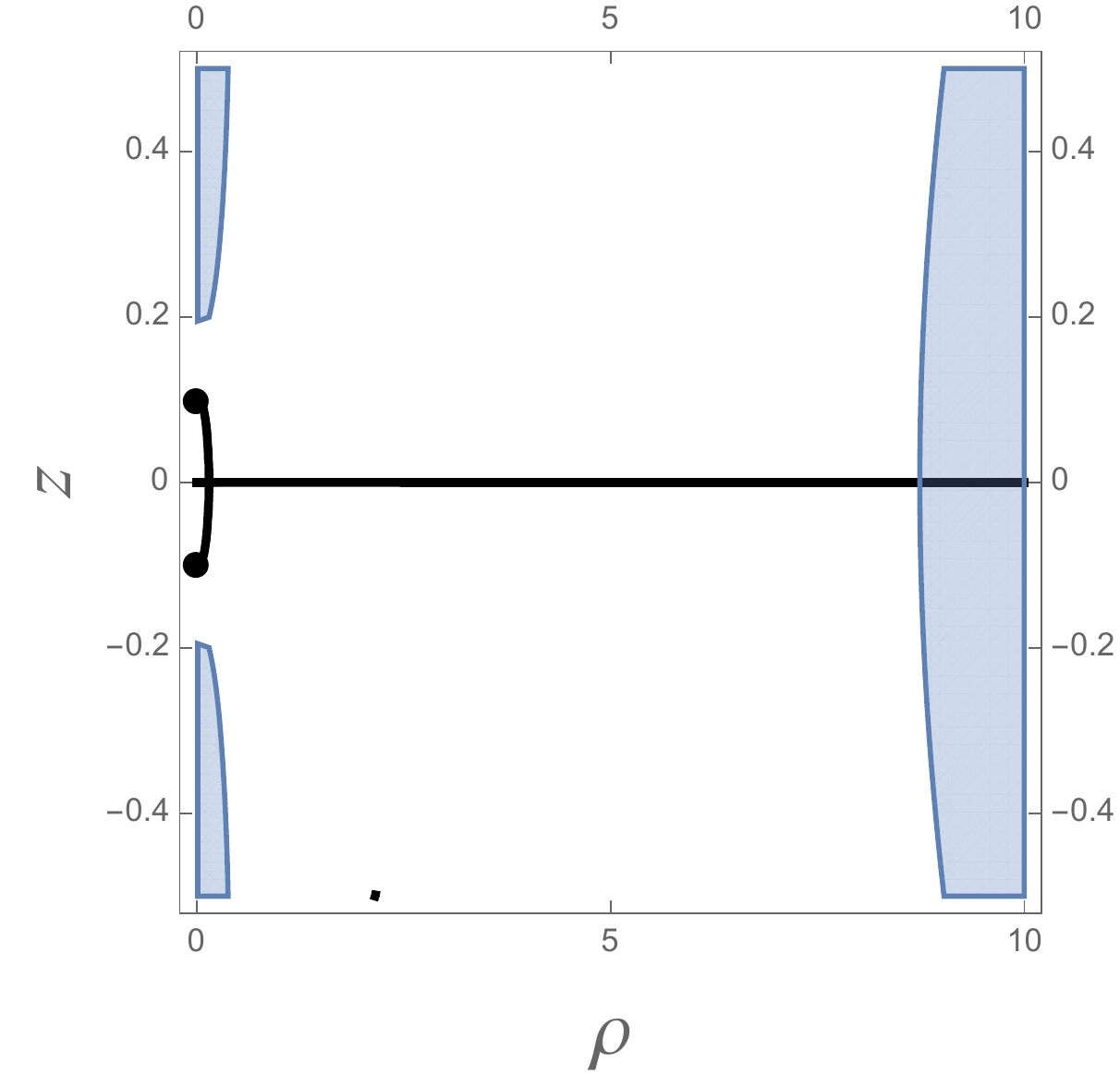}
~~~~
\caption{$M=1, z_1=0.1$, $l=5$
}
\label{fig:a01_4_5}
 \end{minipage} &
 
\begin{minipage}[t]{0.5\hsize}
\centering
\includegraphics[width=6cm]{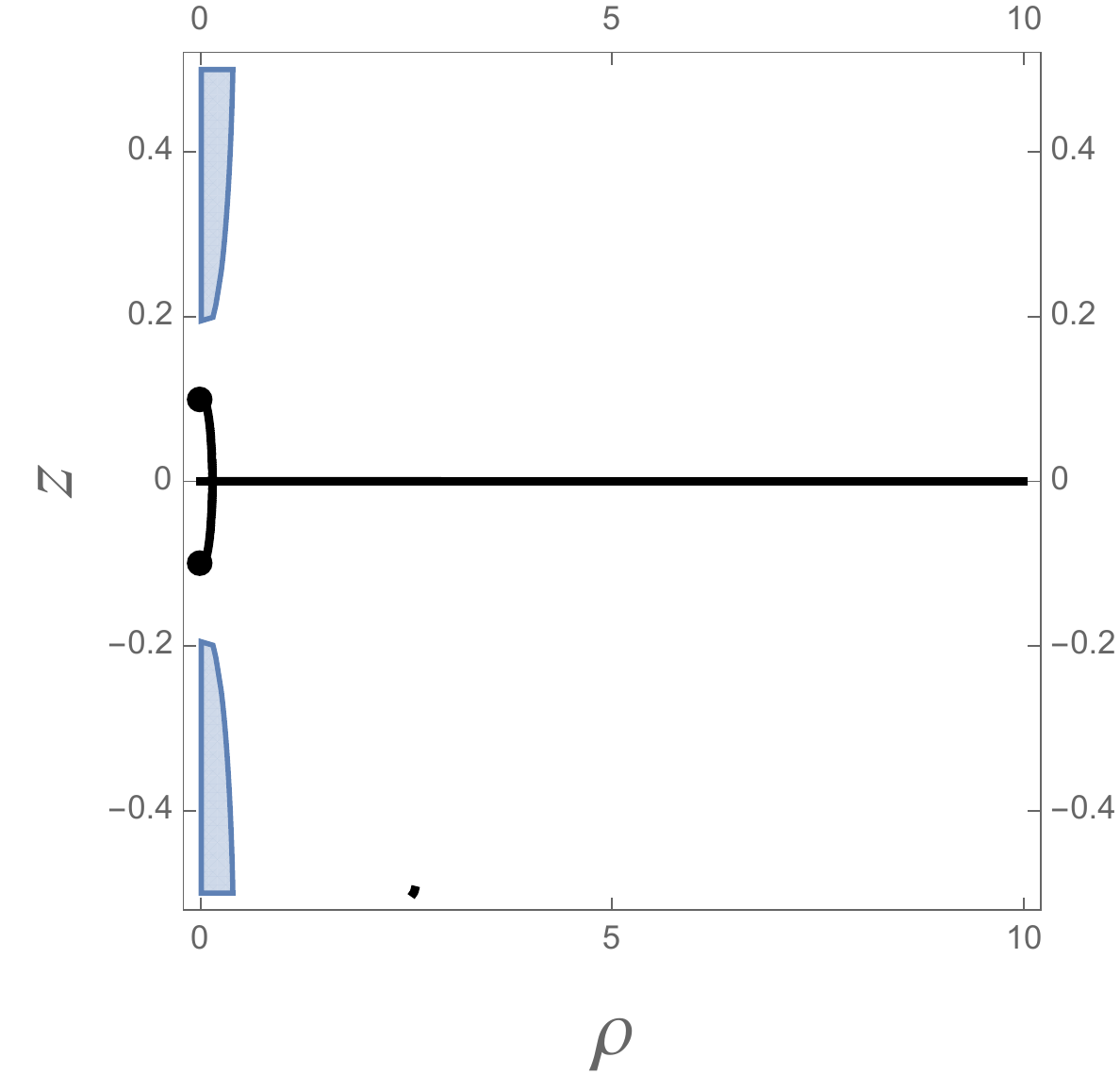}
~~~~
\caption{$M=1$, $z_1=0.1$, $l=10$}
\label{fig:a01_5}
 \end{minipage}

\end{tabular}
\end{figure}

\medskip
Figures~\ref{fig:a01_4}--\ref{fig:a01_5} show the typical shapes of the existence region of stable circular orbits for a sufficiently small separation $z_1\ll M$ in the choice of parameters $(M,z_1)=(1,0.1)$ and $l=1, 4, 5, 10$.
The stable circular orbits exist only on the $z=0$ branch but not on the $z\not=0$ branch, regardless of the size of the extra dimension. 
On the $z=0$ branch, the existence region of the stable circular orbits has only a semi-infinite region $\rho_{\mathrm{out}}<\rho<\infty$, where $\rho_{\mathrm{out}}$ corresponds to the radius of an innermost stable circular orbit.
This implies that the inner portion seen in a large separation case vanishes, while only the outer portion exists.
Furthermore, as is seen from these figures, as $l$ becomes large, 
$\rho_{\mathrm{out}}$ becomes large. 
In the limit of $l\to\infty$, the radius $\rho_{\mathrm{out}}$ goes to infinity, and hence, the stable circular orbits vanish.

\medskip
Finally, we explain the reason why there are stable circular orbits at infinity. 
On the $z=0$ branch in the asymptotic region,
$L_{\varphi 0}^2$, $h_0$, $k_0$
behave as, respectively, 
\begin{eqnarray}
L_{\varphi 0}^2(\rho, 0)&\simeq& 2M\rho,\\
h_0(\rho, 0)
&\simeq& \frac{16M^2}{\rho^6},\\
k_0(\rho, 0)
&\simeq& \frac{8M}{\rho^3} .
\end{eqnarray}
The positivity of these quantities implies that on the $z=0$ branch in the asymptotic region, there always exists an overlap region $\gamma_0\cap D$, where 
we can always find stable circular orbits. 
We can physically interpret that the spacetime effectively behaves as four-dimensional one in the asymptotic region because of the compactification of an extra dimension.

%%%%%%%%%%%%%%%%%%%%%%%%%%%%%%%%%%%%%%%%%%%%
%%%%%%%%%%%%%%%%%%%%%%%%%%%%%%%%%%%%%%%%%%%%
%%%%%%%%%%%%%%%%%%%%%%%%%%%%%%%%%%%%%%%%%%%%

%                                << Summary>>

%%%%%%%%%%%%%%%%%%%%%%%%%%%%%%%%%%%%%%%%%%%%
%%%%%%%%%%%%%%%%%%%%%%%%%%%%%%%%%%%%%%%%%%%%
%%%%%%%%%%%%%%%%%%%%%%%%%%%%%%%%%%%%%%%%%%%%
\section{Summary and discussions}\label{sec:summary}

In this work, we have shown the existence of the stable circular orbits for massive particles around the static squashed Kaluza-Klein black holes with a single horizon and two horizons in the five-dimensional Einstein-Maxwell theory (five-dimensional minimal supergravity), 
 reducing the geodesic motion of particles to a two-dimensional potential problem. 
For a single horizon, we have shown analytically that there always exists an innermost stable circular orbit, whose radius monotonically depends on the size of the extra dimension. 
In the large limit of the extra dimension (to an asymptotically flat black hole), the radius goes to infinity and the stable circular orbits vanish.
This is consistent with the result we have expected. 
Moreover, for two horizons, the existence region of the stable circular orbits depends on the size of an extra dimension and the separation between two black holes. 
For a sufficiently large separation, on two branches, the $z=0$ branch and the $z\not=0$ branch, the stable circular orbits exist regardless of the size of extra dimension, whereas 
for a sufficiently small separation, only on the $z=0$ branch they exist, and the radius of an innermost stable circular orbit monotonically increases with the extra dimension size. 
In particular, for a large separation, on both branches stable circular orbits exist for the small extra dimension, and as $l$ becomes large, the set of the stable circular orbits on the $z=0$ branch separates into two parts (an inner portion and an outer portion), and the outer portion goes to infinity as $l\to \infty$.

\medskip
It is interesting to compare this result of the stable circular orbits in the Kaluza-Klein solution (with a single horizon) with that of the caged black hole solution in Ref.~\cite{Igata:2021wwj}. 
Similarly to the Kaluza-Klein black hole, for the caged black hole, in the asymptotic region, stable circular orbits always exist because of the small extra-dimensional space, whereas in the vicinity of the black hole, they do not exist because the effect of compactification is no longer effective. 
 However, when the size of an extra dimension becomes smaller and smaller, the explicit difference appears. 
As it becomes smaller, in the caged black hole background, the radius of an innermost stable circular orbit becomes larger, but in contrast, in the Kaluza-Klein black hole background it becomes smaller.

\medskip
Moreover, it is also interesting to compare the result of the stable circular orbits in the Kaluza-Klein solution (with two horizons) with that of the Majumdar-Papapetrou solution (with two horizons) in Ref.~\cite{Igata:2020vlx}.
In the Majumdar-Papapetrou background, for a large separation, stable circular orbits exist from the vicinity of the horizons to infinity; for a median one, they appear only in a certain finite region bounded by the innermost stable circular orbit and the outermost stable circular orbit outside the horizons; and for a small one, they do not appear at all. 
The most significant difference appears in the asymptotic region. 
In the asymptotic region of the Kaluza-Klein background, 
stable circular orbits can exist for any parameters, whereas in the asymptotic region of the Majumdar-Papapetrou background, they can only exist for a large separation.

\medskip
In this work, for simplicity we have considered only the motion of particles with $L_\zeta=0$, which implies that particles on the $z=0$ branch have no electric charges in a dimensionally reduced four-dimensional spacetime.
The generalization of our analysis to particles with $L_\zeta\not=0$ may be of physical interest 
because such particles have electric charges from a four-dimensional point of view and hence  are subject to the Coulomb force from black holes. 
This will significantly change the existence region of stable circular orbits. 
This issue deserves further study.

 %%%%%%%%%%%%%%%%%%%%%%%%%%%%%%%%%%%%%%%%%%%%
%%%%%%%%%%%%%%%%%%%%%%%%%%%%%%%%%%%%%%%%%%%%
%%%%%%%%%%%%%%%%%%%%%%%%%%%%%%%%%%%%%%%%%%%%

%                                << acknowledgments>>

%%%%%%%%%%%%%%%%%%%%%%%%%%%%%%%%%%%%%%%%%%%%
%%%%%%%%%%%%%%%%%%%%%%%%%%%%%%%%%%%%%%%%%%%%
%%%%%%%%%%%%%%%%%%%%%%%%%%%%%%%%%%%%%%%%%%%%

\acknowledgments
This work was supported by the Grant-in-Aid for Scientific Research (C) [JSPS KAKENHI Grant Number~17K05452~(S.T.)] and the Grant-in-Aid for Early-Career Scientists [JSPS KAKENHI Grant Number~JP19K14715~(T.I.)] from the Japan Society for the Promotion of Science. S.T. is also supported from Toyota Technological Institute Fund for
Research Promotion (A).

%%%%%%%%%%%%%%%%%%%%%%%%%%%%%%%%%%%%%%%%%%%%
%%%%%%%%%%%%%%%%%%%%%%%%%%%%%%%%%%%%%%%%%%%%
%%%%%%%%%%%%%%%%%%%%%%%%%%%%%%%%%%%%%%%%%%%%

%                                << Appendix >>

%%%%%%%%%%%%%%%%%%%%%%%%%%%%%%%%%%%%%%%%%%%%
%%%%%%%%%%%%%%%%%%%%%%%%%%%%%%%%%%%%%%%%%%%%
%%%%%%%%%%%%%%%%%%%%%%%%%%%%%%%%%%%%%%%%%%%%
\appendix
\section{Static multi--Kaluza-Klein black holes as supersymetric solutions}
Let us begin with supersymmetric solutions in the five-dimensional minimal ungauged supergravity~\cite{Gauntlett:2002nw}, whose bosonic Lagrangian consists of the Einstein-Maxwell theory with a Chern-Simons term.
In the theory, the metric and the gauge potential of the supersymmetric solutions take the form:
\begin{eqnarray}
\label{metric}
ds^2&=&-f^2(dt+\omega)^2+f^{-1}ds_{M}^2,\\
A&=&\frac{\sqrt 3}{2} \left[f(d t+\omega)-\frac KH(d \psi+\chi)-\xi \right]\,,
\end{eqnarray}
where $ds^2_M$ is the Gibbons-Hawking metric,
\begin{eqnarray}
ds^2_M&=&H^{-1}(d\psi+\chi)^2+H(dx^2+dy^2+dz^2), \quad 
\chi=\sum_{i=1}^nh_i\tilde \omega_i,\\
H&=&h_0+\sum_{i=1}^n\frac{h_i}{r_i}, \label{Hdef}
\end{eqnarray}
where $r_i:=|{\bm r}-{\bm r_i}|=\sqrt{(x-x_i)^2+(y-y_i)^2+(z-z_i)^2}$, ${\bm r}:=(x,y,z)$ and ${\bm r}_i:=(x_i,y_i,z_i)$. 
The vectors $\partial/\partial t$ and $\partial/\partial \psi$ are Killing vectors. 
The others are written as
\begin{eqnarray}
f^{-1}&=&H^{-1}K^2+L,\\
\omega&=&\omega_\psi(d\psi+\chi)+\hat \omega,\\
\omega_\psi&=&H^{-2}K^3+\frac{3}{2} H^{-1}KL+M, \\
\hat \omega&=&\sum_{i,j=1}^n\left(h_i m_j+\frac{3}{2}k_il_j\right)\hat\omega_{ij}-\sum_{i=1}^n\left( m_0h_i+\frac{3}{2}l_0k_i \right)\tilde \omega_i,
\\
\xi&=&-\sum_{i=1}^n k_i \tilde{\omega}_i,\\
\\
K&=&\sum_{i=1}^n\frac{k_i}{r_i},\quad L=l_0+\sum_{i=1}^n\frac{l_i}{r_i}, \quad M=m_0+\sum_{i=1}^n\frac{m_i}{r_i},
\end{eqnarray}
and the one-forms $\tilde\omega_i$ and $\hat\omega_{ij}$ on ${\mathbb E}^3$ are determined by $*d\tilde \omega_i=d(1/r_i)$ and $*d\tilde \omega_{ij}=(1/r_i)d(1/r_j)-(1/r_j)d(1/r_i)$. 
Under the choice of the parameters $l_0=1, k_i=m_0=m_i=0\ (i=1,\ldots, n)$, the functions $f^{-1}$, $\omega$ and $\xi$ can be simply written as 
\begin{eqnarray}
f^{-1}&=&L,\\
\omega&=&0,\\
\xi&=&0,
\end{eqnarray}
and the metric is reduced to
\begin{eqnarray}
ds^2&=&-L^{-2}dt^2+L \left[ H^{-1}(d\psi+\chi)^2+H(dx^2+dy^2+dz^2) \right]\\
        &=&-L^{-2}dt^2+L \left[  (H/h_0)^{-1} (d\psi/\sqrt{h_0}+\chi/\sqrt{h_0})^2+(H/h_0) h_0 (dx^2+dy^2+dz^2)\right] \\
        &=&-\bar H^{-2}dt^2+\bar H \left[  \bar V (d\bar \zeta+\bar{\bm \omega})^2+\bar V^{-1}  (d\bar x^2+d\bar y^2+d\bar z^2)\right], 
\end{eqnarray}
where we have replaced as follows:
\begin{eqnarray}
&&L =1+\sum_{i=1}^n\frac{l_i}{r_i} \to \bar H=1+\sum_{i=1}^n\frac{M_i}{\bar r_i},\\
&&H/h_0=1+\sum_{i=1}^n\frac{h_i/h_0}{r_i} \to \bar V^{-1} =1+\sum_{i=1}^n\frac{N_i}{\bar r_i},\\
&&\psi/\sqrt{h_0}\to \zeta,\\
&& \sqrt{h_0} x^i\to \bar x^i,\\
&& \sqrt{h_0} r_i \to \bar r_i=|{\bm x}-{\bm x}_i|,\\
&& \sqrt{h_0}l_i \to M_i,\\
&& h_i/\sqrt{h_0} \to N_i.
\end{eqnarray}
This coincides with the solution in Ref.~\cite{Ishihara:2006iv}.

%%%%%%%%%%%%%%%%%%%%%%%%%%%%%%%%%%%%%%%%%%%%
%%%%%%%%%%%%%%%%%%%%%%%%%%%%%%%%%%%%%%%%%%%%
%%%%%%%%%%%%%%%%%%%%%%%%%%%%%%%%%%%%%%%%%%%%

%                                << Reference>>

%%%%%%%%%%%%%%%%%%%%%%%%%%%%%%%%%%%%%%%%%%%%
%%%%%%%%%%%%%%%%%%%%%%%%%%%%%%%%%%%%%%%%%%%%
%%%%%%%%%%%%%%%%%%%%%%%%%%%%%%%%%%%%%%%%%%%%

\end{document}